\documentstyle[12pt,aasms4]{article}

\begin{document}

\title{The stellar content of the Local Group dwarf galaxy Phoenix\altaffilmark{1}}

\author{D. Mart\'\i nez-Delgado}

\affil{Instituto de Astrof\'\i sica de Canarias, E-38200 La Laguna, Canary Islands, Spain\\ e-mail: ddelgado@ll.iac.es}

\author{C. Gallart}

\affil{Observatories of the Carnegie Institution of Washington, 813 Santa Barbara St., Pasadena, CA 91101, USA \\ e-mail: carme@ociw.edu}
\and

\author{A. Aparicio}

\affil{Instituto de Astrof\'\i sica de Canarias, E-38200 La Laguna, Canary Islands, Spain \\ e-mail: aaj@ll.iac.es}

\altaffiltext{1} {Based on observations made with the 100$''$ du Pont telescope of the Carnegie Institution of Washington at Las Campanas Observatory in Chile.}

\begin{abstract}

We present new deep $VI$ ground-based photometry of the Local Group dwarf galaxy Phoenix. Our results confirm that this galaxy is mainly dominated by red stars, with some {\it blue plume} stars indicating recent (100 Myr old) star formation in the central part of the galaxy. 

We have performed an analysis of the structural parameters of Phoenix based on 
an ESO/SRC scanned plate, in order to search for differentiated components. The elliptical isopleths show a sharp rotation of $\simeq 90 \deg$ of their major axis at radius $r\simeq 115\arcsec$ from the center, suggesting the existence of two components: an inner component facing in the east--west direction,
 which contains all the young stars, and an outer component oriented   north--south, which seems to be  predominantly populated by old stars. These results were then used to obtain the color--magnitude diagrams for three different regions of Phoenix in order to study the variation of the
 properties of its stellar population.

The young population located in the central component of Phoenix shows a clear asymmetry in its distribution, with the younger {\it blue plume}  stars predominantly located in the western half of the central component and the older core helium-burning stars  predominantly situated in the east. This spatial variation could indicate a propagation of star formation across the central component. The H~{\sc i} cloud found at $\sim6\arcmin$ Southwest by Young \& Lo (1997) could have been involved in this process. We also find a decreasing gradient in the density of intermediate-age population with the galactocentric radius, based on the number of stars populating the red clump in the 
color--magnitude diagram. Since no metallicity gradient is apparent, this indicates the presence of a substantial intermediate-age population in the central region of Phoenix that would be less abundant or absent in its outer regions. This result is also consistent with the gradient found in the number of horizontal branch stars, whose frequency relative to red giant branch stars increases towards the outer part of the galaxy. These results, together with those of our morphological study, suggest the existence of an old, metal-poor population with a spheroidal distribution surrounding the younger inner component of Phoenix. This two-component structure may resemble the halo--disk structure observed in spirals, although more data, in particular on kinematics, are necessary to confirm this.

We have estimated the average star formation rate for the last 1 Gyr  and for the age interval 1--15 Gyr from the number of blue and red giant branch and asymptotic giant
branch stars observed in the color--magnitude diagram. For the central region, the average past star
formation rate is very similar to that for the last 1 Gyr. The recent star formation rate of Phoenix is also comparable to that displayed by typical dIrr galaxies except perhaps for the fact that it lacks any strong very recent burst as exhibited by galaxies such as Sextans A or NGC 6822. The area-normalized star formation rate for the central region of Phoenix is in the range obtained by Hunter \& Gallagher (1986) for their sample of dIrr galaxies.

We have determined a distance modulus for Phoenix of $(m-M)_0=23.0\pm 0.1$ using the tip of the red-giant branch as a distance indicator. We find four short-period variable candidates from our photometry that might be anomalous
Cepheids or W Vir stars. Finally, it is very unlikely that Phoenix has globular clusters, as is expected for a galaxy with such a faint absolute magnitude.
\end{abstract}

\keywords{galaxies:dwarf --- galaxies:Local Group --- galaxies:stellar content --- galaxies:structure--- techniques: photometry}

\section{Introduction} \label{introduccion}

In generic hierarchical clustering scenarios for galaxy formation, such as cold dark matter dominated cosmologies (White \& Rees 1978; Blumenthal {\it et al.} 1984; Dekel \& Silk 1986), dwarf galaxies should have formed {\it prior} to the epoch of giant galaxy formation and would be the building blocks of larger galaxies. The dwarf galaxies observed today would be surviving objects that have not merged with larger galaxies, and their underlying structure could provide important clues about the process of dwarf galaxy formation at high redshifts.   Local Group dwarf galaxies offer the only opportunity to study their evolution in great detail through the observation of their resolved stellar populations, which  provide fossil records of their star formation history (SFH) and chemical enrichment. In addition, detailed observations of their two-dimensional structure, designed to search for gradients in the stellar populations, permit the determination of the internal {\it structure} of these systems and   provide direct input for detailed models of dwarf galaxy formation.  In this paper, we present observations of the Phoenix dwarf galaxy, providing a full spatial coverage of the system that allows us to study the stellar population gradients of the galaxy. 

The Phoenix dwarf galaxy was discovered by Schuster \& West (1976) from visual inspection of a {\it B} plate of the ESO survey. The first photographic photometry  by Canterna \& Flowers (1977), which showed the presence of some young blue stars, suggested this low surface brightness object was a distant dwarf irregular galaxy. Deeper color--magnitude diagrams (CMDs) obtained by Ortolani \& Gratton (1988) brought Phoenix into the Local Group, at a distance of about 500 kpc, and showed that it is a dwarf galaxy dominated by an old, metal-poor population, with a small number of young stars. These authors proposed that Phoenix belongs to an intermediate class between typical irregular galaxies and dwarf spheroidals, because it shares properties of both types of galaxies. Similar conclusions were reached by Van de Rydt, Demers \& Kunkel (1991, VDK hereafter) in their $BVI$ photometry study of the stellar content of Phoenix. They found Phoenix to be still closer ($\sim$ 400 kpc) to the Milky Way and argued that the few blue stars observed in its CMD can be explained as a burst of star formation produced less than 150 Myr ago. The possible presence of upper asymptotic giant branch (AGB) stars also suggested the existence of an intermediate-age population and hence
a complex star formation history for this galaxy.

Phoenix is situated on the west edge of the Magellanic Stream, whose H~{\sc i} structure is complex and multivalued near the  position of Phoenix 
(Morras 1985). Therefore, although several H~{\sc i} surveys have detected small amounts of gas near Phoenix, the current lack of an optical velocity for the galaxy means that an association between the H~{\sc i} and Phoenix is speculative. Carignan {\it et al.} (1991) detected H~{\sc i} emission at $V_{\sun}$=56 km s$^{-1}$, clearly separated from a much larger scale component at $\sim$ 120 km s$^{-1}$ associated with the Magellanic Stream previously detected by Morras \& Bajaja (1986). Oosterloo, Da Costa \& Staveley-Smith (1996) reported that the H~{\sc i} detected by Carignan {\it et al.} (1991) is part of a large H~{\sc i} complex extending over
more than one degree and is thus  not likely to be associated with Phoenix. These authors detected a smaller cloud at $V_{\sun}= -35$ km s$^{-1}$, situated
at 6$\arcmin$ SW from the center of the galaxy, which could be associated
with Phoenix. Recent VLA observations by Young \& Lo (1997) ahve
 confirmed the existence of this cloud (cloud A) close to the optical body of the galaxy, which forms a curved structure that wraps around its SW border. Furthermore, they confirmed there is no H~{\sc i} coincident with the optical body and detected additional H~{\sc i} about 10$\arcmin$ south of it, with $V_{\sun}$=55 km s$^{-1}$. Hence, it is very important to measure the optical radial velocity of Phoenix in order to confirm any possible association with these H~{\sc i} clouds. It that were the
case, it would be possible to obtain the $M_{\rm H I}/L_{B}$ relation, which is an important parameter necessary for understanding the evolutionary status of Phoenix. 

In this paper, we discuss the stellar content of Phoenix in the light
of new ground-based $VI$ CCD photometry. Brief progress reports of
this study were given by Mart\' \i nez-Delgado, Gallart \& Aparicio (1998),
Mart\' \i nez-Delgado \& Aparicio (1998a) and Mart\' \i nez-Delgado {\it
et al.} (1999). In Sec. \ref{observacion} we present the observations, data reduction, profile-fitting photometry and artificial-star tests. Sec. \ref{structure}  discusses  the spatial structure of Phoenix on the basis of digitized photographic material. In Sec. \ref{cmd} we present the CMD of Phoenix from our CCD photometry. In Sec. \ref{distance} we calculate its distance using the tip of the red giant branch (TRGB). Sec. \ref{gradient} investigates the spatial variation of the different stellar components of Phoenix. In Sec. \ref{variable} we study the variable stars. In Sec. \ref{globular} we
discuss the possible existence of globular clusters in Phoenix. Finally, in Sec. \ref{conclusion} the results presented throughout the paper are summarized.

\section{Observations and data reduction } \label{observacion}

Observations of the Phoenix dwarf galaxy were obtained in  1997 February at the
du Pont 100$''$ telescope at Las Campanas Observatory (LCO) under photometric conditions. The detector used was a 2048 $\times$ 2048 thinned Tektronix chip, which provided a pixel size of $0.296\arcsec$ and a total field of $8\arcmin$.
	
	The field was centered  $4\arcmin$ north west from the center of Phoenix with the purpose of including  the external regions of the galaxy and avoiding the bright star situated close to its border. Images in $V$ and $I$ bands were obtained during nine consecutive nights as part of a strategy to
search for Cepheid variable stars. The best seeing images ($<1.0\arcsec$) from this campaign were selected for this study. The total integration times were 2100 s in $V$ and 2700 s in $I$.  In addition, 1200-s exposures in both $V$ and $I$ of a comparison field situated $7.5\arcmin$ 
east of the center of Phoenix were taken with the purpose of studying the
foreground contamination. The journal of observations for the frames selected for this study is given in Table~1. A $B$ image was taken during a different observing run at LCO to obtain the color image of the central region of Phoenix shown in Figure \ref{color} from the $B$, $V$ and $I$ images using the IRAF task RGBSUN.

	The IRAF CCDRED package was used to correct the data for overscan and flatfield,  using bias and high signal-to-noise sky flats taken  each night.
The $I$ images were corrected from the usual interference pattern using a fringe frame obtained from the median of a set of nearly blank field images acquired during the observing run. This fringe image was then suitably scaled and removed from the $I$ images.

\subsection{Photometry} \label{photo}

	The photometry of the stars in Phoenix has been obtained using the
ALLFRAME software (Stetson 1994), kindly made available to us by Dr. Stetson. A medianed frame was obtained from the combination of the seven $V$ and 
seven
 $I$ images selected for this study. The DAOPHOT II/ALLSTAR package was used to obtain a complete list of stars on the medianed image, using three FIND/ALLSTAR/SUBSTAR passes. 
This list was given as input to ALLFRAME to perform the photometry in 14
individual frames. Only stars measured in at least
three images were retained. PSFs computed in each individual frame were used in this process. We adopted a Moffat function with $\beta$=2.5 plus a table of residuals which vary quadratically with position in the frame.

        DAOMASTER was used to obtain the instrumental $V$ and $I$ magnitudes as the error-weighted average of the magnitudes for each star in the individual frames, selecting only stars with $\sigma_{V} < 0.2$ and $\sigma_{I} < 0.2$.  Next, we paired the $V$ and $I$ list to obtain $(V-I)$ color indexes. The final number of measured stars is 6530. The $\sigma$ standard errors of the mean of these stars are plotted in Figure~\ref{error}. 

        Atmospheric extinction corrections for each night and the photometric transformation to the Johnson--Cousins standard system were derived from  observations during the campaign  of a total of 156 measures in each band $V$ and $I$ of 37 standard stars from the list of Landolt (1992). The final equations used to transform our instrumental magnitudes into the Johnson--Cousins magnitudes are

 	$$ (V-v)= 24.517-0.027(V-I)$$
 
	$$ (I-i)= 24.340-0.029(V-I)$$

The photometric conditions during the campaign were stable and produce
very small zero point errors for the photometric transformation: $\pm 0.006$ mag in $V$ and $\pm 0.008$  mag in $I$.
The standard errors in the extinction were always smaller than 0.012 in both 
$V$ and $I$. Aperture corrections were obtained from a large number of isolated stars in one $V$ and one $I$ image of Phoenix. These are accurate to $\pm$0.010 mag in $V$ and $\pm$0.009 mag in $I$. Taking all these uncertainties into account, we estimate a total error in the photometry zero point of 0.02 mag in both $V$ and $I$.

\subsection{Artificial Stars Tests}\label{crowding} 

Artificial star tests have been performed in a way similar to that 
described in 
Aparicio \& Gallart (1995) and Gallart {\it et al.} (1996a) to obtain 
the necessary information about the observational effects present 
in the photometry. In short, a number of artificial stars of known 
magnitudes and colors are injected in the original frames (to produce what we call {\it synthetic frames})  using the ADDSTAR algorithm of DAOPHOT~II (Stetson 1993), and the 
photometry is done again following the same procedure 
used to obtain the photometry for the original frames. This process 
is repeated as many times as necessary to test a large 
number of artificial stars (25000 in this particular case). 
The injected and recovered magnitudes of the artificial stars, 
together with the information from those that  
were lost, provide the necessary information for all the 
observational effects present in the photometry (see 
Aparicio \& Gallart 1995 for a thorough discussion of these 
effects).

 The distribution of the stars in the CMD has been chosen in a way similar to
that of  Mart\'\i nez-Delgado \& Aparicio (1998b), using a synthetic CMD constructed with constant star formation rate (SFR; from 15 to 0.01 Gyr) and metallicity, $Z$, linearly increasing with time from 0 to 0.002. The main difference from the former paper is that  the procedure here has been adapted to the current data reduction performed with ALLFRAME, and is a longer and more complicated procedure. 

Because each artificial star trial requires a large amount of CPU time, we have optimized the distribution of artificial stars in the real frames by compromising between adding the largest possible number of stars to each frame and retaining its original 
crowding characteristics. All we require for the artificial-star tests to 
represent the actual crowding and other observational effects of our images is that the artificial stars do not interact with each other, {\it i.e.,} 
that the recovered magnitudes are  affected only by the stars in the 
galaxy plus any other artifacts in the image possibly affecting the photometry. 

We constructed a grid with $(x,y)$ coordinates of the artificial stars equally spaced within it. This grid is displaced randomly for each test to assure a sampling of the observational effects over the frame. The distance between the nodes of the grid is such that the effect on each artificial star produced by the wings of its artificial neighbor is negligible. A distance of (2$\times$PSF radius +1) fulfils this requirement, because the wings of the artificial stars extend as far as the PSF radius. With a PSF radius of 12 pixels in our frames, the artificial stars have been distributed 25 pixels apart.
With this spacing, we can add approximately 6300 artificial stars 
per trial to each frame, and the number of runs of artificial-star tests 
reduces to four.  

  Figure \ref{artificial} shows the injected magnitudes and the resulting recovered magnitudes for the artificial stars used in this analysis of the observational effects in Phoenix.

\section {The spatial structure of Phoenix} \label{structure}

The study of the structural parameters of dwarf galaxies can provide valuable information about their formation and evolution, as well as the possible relationships between different dwarf galaxy types. An open debate exists 
concerning the possible secular evolution of dIrr to dEs based on the morphological similarities between both types of galaxies, such as the lack of a conspicuous nucleus or the fact that their radial profiles can be fitted by an exponential law in both cases (Lin \& Faber 1983; Ferguson \& Binggeli 1994). 
Phoenix is a unique target for approaching this issue because it seems to be a nearby prototype  of  dIrr/dE. 

 VDK obtained some details about the structure of Phoenix from a scanned ESO/SRC IIIaJ plate of the galaxy. They found that
Phoenix follows an exponential surface brightness profile, which reaches the zero level at $\sim$ 8.7$\arcmin$ along the major axis but 
 does not fit a King model. They concluded that Phoenix is a rather isolated galaxy which has never been tidally affected by the Milky Way. Apart from this, there is no other detailed study of the structural parameters of Phoenix in the literature. 

	We have performed an analysis of the structural parameters of Phoenix using a method similar to that described by Irwin \& Hatzidimitriou (1995, IH hereafter). Our results are presented in this section.

 \subsection{Photographic plate material and star list}\label{plate}

It is not possible to derive the global structural parameters of Phoenix from
our CCD image because its field is not large enough to cover the whole galaxy. We used a portion of a plate from the UK Schmidt Telescope (UKST) survey that Dr. Irwin kindly digitized for us. The exposure time for this plate was 60 minutes in IIIaJ emulsion with a GG395 filter, reaching a limiting magnitude of $B\sim 22$. 
Figure \ref{plate2} shows the scanned region of this plate used in
this study. This region covers a  35$\arcmin$ $\times$ 35$\arcmin$ field 
centered on Phoenix and was digitized at a pixel sampling of 1$\arcsec$ using the Automatic Plate Measuring System (APM, Kibblewhite {\it et al.} 1984) at Cambridge. Because of we are using a central portion of a larger plate ($6\arcdeg
\times 6\arcdeg$), the resulting image is quite uniform and there are no systematic magnitude differences caused by optical vignetting.

DAOPHOT II/ ALLSTAR software (Stetson 1993) was used to make a catalog of the objects detected in the plate. A preliminary list of stars was obtained after choosing an adequate FIND threshold. This list was given as input to ALLSTAR, which performed a profile-fitting photometry using the PSF of the plate. A total of 11032 stars were measured.

\subsection{Isopleth maps of Phoenix }\label{isopleth}

	 Isopleth maps of the distribution of stars in a resolved galaxy is a means of analyzing its two-dimensional structure. The density maps of this study have been obtained by means of a kernel method similar to that discussed by Alfaro, Delgado \& Cabrera-Ca\~no (1992). In short, a map with the position of the stars was convolved with a two-dimensional Cauchy kernel with a smoothing parameter of $h_{x}=h_{y}=25\arcsec$. The result of this process is a smooth map of the spatial stellar density of the galaxy, which gives a visual description of its morphology. However, the density maps derived from our star list are affected by two factors: i) contamination by foreground Galactic stars and background  galaxies, and ii) crowding effects in the inner part of the galaxy. Both effects are difficult to estimate, and in the case of background contamination are unreliable at fainter magnitudes.  For this reason we adopted an approach similar to that of IH, using all the raw stellar images detected to construct the isopleth map  and correcting {\it a posteriori} the obtained radial profile for crowding effects and foreground contamination (see Sec. \ref{profile}).

Figure \ref{isopleth_fig} shows the isopleth map obtained for Phoenix. The structural parameters can be determined by fitting ellipses to each isodensity contour. This provides a reliable means of deriving the center, ellipticity ($\epsilon=1-b/a$) and position angle (PA) of the number distribution of stars. The parameters obtained by this procedure do not exhibit significant variations through the galaxy, except at  $r\sim 115\arcsec$ from the center, where a sharp rotation of $\sim 90\arcdeg$ in the orientation of the semi-major axes of the ellipses is observed. This  change suggests the existence of two components in Phoenix: an inner component oriented   east--west; and an outer component which is more extended and clearly aligned north--south. Interestingly, the inner component contains the majority of the galaxy's young stars, with a flattened distribution fitting its shape quite well (see Sec. \ref{bp}). This, together with the clearly different orientation from the outer component
could indicate that the galaxy has a disk--halo structure similar to that of
bigger, spiral galaxies. However, data on the kinematics are necessary before
definitely stating that this is the case. Table~2 lists the mean of the center (in pixels), ellipticity and PAs for both components. The outer component was previously detected by VDK in their study of the morphology of Phoenix. They found a similar ellipticity  ($\epsilon \sim 0.3$) and orientation as
that given here for this component.

\subsection{Radial profile and model fitting }\label{profile}

The radial density profile of Phoenix can be derived from the original star
list described in Sec. \ref{plate} by counting stars within suitably oriented elliptical annuli centered on the galaxy center. We will  adopt the radial profile obtained using the values of the center, ellipticity and PA for the outer component (Table 2) because it allows a more accurate study of the outer regions of Phoenix (see Sec. 3.4).

As mentioned in Sec. \ref{isopleth}, this raw radial profile must be corrected 
for observational effects and background contamination (which includes Galactic stars and background galaxies). Evaluation of the background contamination  is critical to determining the outer regions of the galaxy and its actual extension ({\it i.e.,} its tidal radius). In our case, the contamination by background galaxies is strong due to the presence of a galaxy cluster in the
field of the scanned region.  To achieve this correction, we adopted the median of the value of the number density at large radii ($r\sim 23 \arcmin$) as the best estimate of the true background level ($d_{\rm bck}$). This value, $d_{\rm bck}=8.0 \pm 0.4 (\arcmin)^{-2}$  was subtracted from the raw radial profile of Phoenix.

Crowding corrections were obtained from artificial-star tests in the digitized
plate, in a similar way as for the CCD images (see Sec. 2.2). A total number of 120000 stars covering the complete range of instrumental magnitudes of the plate were added to the image in small groups of 1200 stars. The resulting frames were processed to obtain the magnitudes of the artificial stars affected by the observational effects, and build a {\it crowding-trial} table (see Aparicio \& Gallart 1995) for the plate. This table was used to derive the crowding factor, $\Lambda$,  in each elliptical annulus. The crowding factor is defined as $\Lambda=N_{\rm rc}/N_{\rm in}$, where $N_{\rm rc}$ is the number of recovered stars and $N_{\rm in}$ is the number of injected stars in each annulus. These  $\Lambda$ values were then used to correct the star counts in each elliptical annulus in order to obtain the crowding-corrected radial profile of Phoenix plotted in Figure 5.

Fitting  models to the radial profiles of dwarf galaxies like Phoenix can provide information about whether they have always been isolated or
whether, on the contrary, they have been tidally truncated in a past encounter with the Milky Way. Two basic models have been extensively applied to study the radial profiles of dwarf galaxies: a) the King model (King 1962), which assumes that the system is isothermal with a Maxwellian distribution of velocities truncated by the presence of a dominant external mass (in our case, the Milky Way); and b) the exponential law, which is based on empirical evidence that  exponential profiles  it galactic disks and  Irr galaxies very well (Faber \& Lin, 1983).
In the case of Phoenix, the existence of the inner and outer structural components probably makes it unrealistic to try fitting a model profile at all distances. Furthermore, since the tidal truncation effects, if present, should
be much more evident in the outer component ($a>112\arcsec$) a King fitting has
been tried for it. We adopt the simple method of King (1962) rather than determine  the tidal and core radii simultaneously. For the outer part, the stellar surface density can be expressed as

$$d(r)= d_{1} (1/r-1/r_{\rm t})^{2}$$,

\noindent where $d_{1}$ is a constant and $r_{\rm t}$ is the tidal radius. This relation
differs only slightly from the exact King model formula given by King (1962; formula 14) in the outer regions of the galaxy. The value of $r_{\rm t}$ can be 
 obtained from the
slope and zero point of a least-square linear fit of a plot of $d^{1/2}$ against $1/r$, as  is shown in Figure \ref{king}. This yields  $r_{\rm t}=15\arcmin.8_{-2\arcmin.8}^{+4\arcmin.3}$, where the errors have been estimated after varying $d_{\rm bck}$ by $\pm 1\sigma$. This is the main source of error in the determination of $r_{\rm t}$ and means that the angular size (tidal diameter) of Phoenix is slightly larger than that obtained in the previous study by VDK ($r_{\rm t}=8.7 \arcmin$). This result is compatible with our CCD observations, where an evident galaxy stellar population is detected at a large galactocentric radius ($\sim 8 \arcmin$). It
also shows that the outer component of Phoenix can be fitting by a King model
and might therefore have been affected by the gravitational field of the Milky Way. Knowledge of the radial velocity of the galaxy could help in constraining the orbit of Phoenix, and therefore give clues about possible past encounters with the Milky Way.

\section{The Color--Magnitude Diagram of Phoenix } \label{cmd}

Figure \ref{cmphoenix} shows the CMD of Phoenix. It has features in common with the CMDs of both dE and dIrr galaxies (Fornax: Stetson, Hesser \& Smecker-Hane 1998; Pegasus: Aparicio {\it et al.} 1997a; Antlia: Aparicio {\it et al.} 1997c), and displays a number of noteworthy stellar populations that provide first insights into the star formation history of this galaxy.

The most prominent feature is the red tangle (see Aparicio \& Gallart 1994), which corresponds to the red giant branch (RGB) and asymptotic giant branch (AGB) loci. This shows, in agreement with previous results, that this galaxy is dominated by red stars (Ortolani \& Gratton 1988; VDK). The color of the RGB stars indicates a low metallicity ([Fe/H]=--1.36; see Sec. \ref{metallicity}) and its dispersion may indicate the existence of an abundance range in Phoenix, although it could also be accounted for by dispersion in age only (see Sec.~\ref{metallicity}). A number of red stars are observed above the tip of the RGB (TRGB), that could be intermediate-age AGB stars. However, the CMD of the comparison field shows a large number of foreground stars in this region of the diagram. The existence of this upper-AGB population is discussed in detail in Sec. \ref{agb}. 

A {\it blue-plume} (BP) composed of young main sequence (MS) stars and/or blue loops is observed at 21 mag $< I <$ 24 mag and $-0.5 < (V-I) < +0.2$,  indicating that recent star formation has taken place in  Phoenix. In addition to this young population, a significant number of stars are observed at $I\sim 22$ mag and $(V-I)\sim 0.8$, close to merging with the RGB locus. These stars are probably core helium-burning intermediate-mass stars with ages $\sim$ 0.8 Gyr (see Sec. \ref{bp}). Some of the stars observed in the strip $0.9<(V-I)<1.1$ and 21 mag $<I<$ 18 mag may correspond to the reddest extreme of the core helium-burning phase of younger stars. 

One of the most striking features is the presence of a well-populated red clump  (RC) at $I\sim 22.5$ mag, $(V-I)\sim 1.0$, and a blue-extended horizontal branch (HB) extending down to $I\sim 23.5$ mag, $(V-I)\sim 0.2$. The RC appears as a widening of the RGB and is the locus of an old to intermediate-age core helium-burning population. The fact that it extends about 1 magnitude in luminosity and  is well-populated indicates that, at least in its central part (see Sec.~\ref{redclump}), the intermediate-age stars make a substantial contribution to the stellar population of Phoenix  (it can be compared with Leo~I: Gallart {\it et al.} 1999a). The blue-extended HB is similar to those observed in other dEs (Tucana: Seitzer {\it et al.} 1998; Fornax: Stetson {\it et al.} 1998). Because its position is very close to our limiting magnitude, we cannot characterize it properly, and  its existence could even be questioned. Nevertheless, its location coincides with that expected for a galaxy at the  distance of Phoenix and it appears horizontal in a $[V,(V-I)]$ CMD. Furthermore, it has also been observed in an independent study of this galaxy (Held, Saviane \& Momany 1999). In the absence of RR Lyrae variable observations, the HB provides the first evidence of an old ($>$
10 Gyr), low-metallicity stellar population in Phoenix.  

\section{The distance to Phoenix} \label{distance}

The distance to Phoenix has been obtained from the absolute $I$ magnitude of the TRGB ($I_{\rm TRGB}$), which has been proven to be an excellent distance estimator for nearby resolved galaxies (Lee {\it et al.} 1993). This point marks the core helium flash of low-mass stars, which occurs at a nearly constant bolometric magnitude in stellar populations with ages in the range 2--15 Gyr and metal abundances in the range $-2.2$ dex $ < [{\rm Fe/H}] < -0.7$ dex.

The position of the TRGB in the CMD of Phoenix has been obtained from the convolution of the $I$ luminosity function with an edge detector, in this case a Sobel filter of kernel (--1,--2, 0.+1,+2; Madore \& Freedman 1995). Figure \ref{tip} shows the $I$ luminosity function (solid line) and the output of this Sobel Filter (dotted line). This produces a sharp peak at the position of the tip, situated at $I_{\rm TRGB}=19.00\pm 0.07$ mag. The error has been estimated from the width of the peak at half its height. We adopted the extinction value for Phoenix of $E(B-V) = 0.02$ from the reddening map of our Galaxy by Burstein \& Heiles (1982). Using the extinction law of Cardelli {\it et al.} (1989), this value leads to $A_{V}=0.06$ and  $A_{I}=0.04$. The resulting dereddened value for the I magnitude of the TRGB is $I_{\rm TRGB, 0}=18.96\pm 0.07$.

The distance modulus is derived using the equation (Lee {\it et al.} 1993):
\begin{equation}
(m-M)_0 = I_{\rm TRGB, 0}+BC_I-M_{\rm bol,TRGB}. 
\end{equation}

The bolometric correction ($BC_{I}$) was computed from the color index of the TRGB $(V-I)_{\rm TRGB}$, using the calibration of Da Costa \& Armandroff (1990)

\begin{equation}
BC_I=0.881-0.243(V-I)_{\rm TRGB,0}.
\end{equation}

A value of $(V-I)_{\rm TRGB}=1.66$ was estimated  from the median color value of 65 stars with 18.75 mag $<I<$ 19.25 mag, which leads to a dereddened color of $(V-I)_{\rm TRGB,0}$=1.64 and to $BC_{I}=0.48$. 

The bolometric magnitude of the TRGB was calculated from the relation (Da Costa \& Armandroff 1990):
\begin{equation}
M_{\rm bol,TRGB}=-0.19[{\rm Fe/H}]-3.81.
\end{equation}

The [Fe/H] value can be estimated from the calibration between metallicity and color of RGB stars at $M_{I,0}=-3.5 $ mag based on old stellar populations in Galactic globular clusters (Da Costa \& Armandroff 1990; Lee {\it et al.} 1993):
\begin{equation}
[{\rm Fe/H}]=-12.64+12.61(V-I)_{-3.5,0}-3.33(V-I)^2_{-3.5,0},
\end{equation}
where $(V-I)_{-3.5,0}$ is the color index of the RGB measured at $M_{I0}=-3.5$ mag. This results in $(V-I)_{\rm -3.5,0}=1.45$, which leads to $[{\rm Fe/H}]=-1.37$ and
 $(m-M)_{0}=23.0 \pm 0.1$ or $d=398 \pm$ 18 kpc. This distance is in good agreement with the mean distance obtained by VDK ($d=417$ kpc).

\section{ Gradients in the stellar population} \label{gradient}

Recent studies have shown that many Local Group dIrr galaxies display significant spatial variations in the morphology of their CMDs (Sextans A: Aparicio {\it et al.} 1987; Dohm-Palmer {\it et al.} 1997; WLM: Minniti \& Zilstra 1996; Leo A: Tolstoy 1996; IC 1613: Hodge {\it et al.} 1991; Antlia: Aparicio {\it et al.} 1997c; Van Dyk, Puche \& Wong 1998), suggesting a varying  SFH across the galaxy. The  stars responsible for the most evident changes in the CMD morphology of these galaxies are young, and therefore the differences in the CMDs across the galaxies are due only to differences in the SFH during the last few hundred million years. In a few cases, a gradient in the old and/or intermediate-age population has also been discussed for the case of dE galaxies (And I: Da Costa et al. 1996; NGC~147: Han et al. 1997) using {\it HST} data. The wide field covered by our CCD images and the deep photometry in a close galaxy like Phoenix allow us to search for gradients in the properties of the stellar component  of any age as a function of the galactocentric radius. This variation can yield important clues which constrain the models of dwarf galaxy formation.  

We have divided the galaxy into three different regions, chosen according to  the structure of Phoenix as described in Sec. \ref{isopleth}: i) an inner elliptical region (region A: 0 pc $< a_{d} \leq$ 229 pc) oriented   east--west,
 which is basically the central component of Phoenix (where $a_{d}$ is the major semiaxis of the central component); and ii) two outer elliptical regions (region B:  229 pc$ < a_{h} \leq$ 457 pc , and region C: $a_{h}>$ 457 pc) both oriented   north--south, that match the outer component of the galaxy. The boundary of these three
regions is shown in Figure \ref{image1}, overplotting the $V$ image of Phoenix.  Figure \ref{3cmd} shows the CMDs for these three regions. At first glance it may be said they display a significant gradient in the stellar population. However, observational effects may play an important role in explaining variations in the CMD morphology (see Mart\' \i nez-Delgado \& Aparicio 1998b). In this section, we will discuss the spatial variation of the different stellar components of Phoenix,  and the role observational effects play in this variation.

\subsection {The young population} \label{bp}

Figure \ref{3cmd}a shows that the BP of young stars is only observed in the CMD of the galaxy's central region. This central concentration of young stars was known in previous studies (Canterna \& Flowers
1977; Ortolani \& Gratton 1988). These stars appear blue in Figure \ref{color}. Their distribution throughout the galaxy is better seen in Figure \ref{mapa}, which displays the position of the BP stars selected from the CMD with  21 mag
$ < I <$ 24 mag and $ -0.5 < (V-I) < 0$ (solid circles).  Interestingly, that distribution matches the shape and the same east--west orientation of the central component found in Sec. \ref{isopleth} (see Figure \ref{isopleth_fig}).

In addition to the  BP stars, several possible core helium burning stars (for simplicity, we will denote these stars HeB in this paper) are also observed in the CMD of the central part of the galaxy (Figure \ref{3cmd}a). This population is better seen in the $[V,(V-I)]$ CMD of the central part of Phoenix (Figure \ref{vcmd}). Their distribution (selected from our photometry with 21.5
mag $<V<$ 22.6 mag and $0.20 < (V-I) <0.98$) is shown in Figure \ref{mapa} (solid triangles). As in the case of the BP stars, they are predominantly located in the center of the galaxy. Unfortunately, our comparison field is not deep enough for a reliable foreground correction of this CMD region. However,
a comparison between the CMDs in the
regions A (Figure \ref{3cmd}a) and C (Figure \ref{3cmd}c) indicates
that many of the HeB candidates are not foreground stars.
Since the surface of region C is seven times larger than that of region A, a larger number of stars in the HeB region should be observed in region C than
in region A, while the opposite is observed.

The location of these stars in the CMD coincides quite well with the position of the intermediate-mass stars in the reddest extent of their HeB phase. This is shown in Figure \ref{vcmd}, in which a theoretical isochrone from the Padua library (Bertelli {\it et al.} 1994) for $Z=0.001$ and age 800 Myr has been overplotted. The lack of information about the metal abundance of the young population of Phoenix prevents us from obtaining an accurate estimate for the age of these stars. Assuming a metal abundance comparable to the maximum value estimated from the RGB stars ($[{\rm Fe/H}]=-1.0$ or $Z=0.001$ if no $\alpha$-enhancement is considered (see Sec. \ref{metallicity}), we calculate an age range of 400--800 Myr for these stars. This population could be younger (300--500 Myr) if it is more metal-rich ($Z=0.004$).

 A significant number of BP stars are clustered in an association situated on the western border of the central region.  This association was previously noticed by Canterna \& Flowers (1977) and has been classified as an association in the literature (Hodge 1998) and corresponds to the latest star formation event in
the galaxy. To estimate its age, we obtained the CMD of a circular region of $r= 52 $ pc centered on the association. We decontaminated the region of the contribution of stars belonging to the general field of Phoenix using the CMD of the region situated in an outer annulus ( 52 pc $<r \leq $104 pc) around the association. Figure \ref{ob} shows the field-subtracted CMD of this association (open circles). A clear MS is observed, although its color dispersion might be due to the severe observational errors rather than to a dispersion in age. Differential reddening is not expected to play a significant role either (see Sec. \ref{recent}). The clump of stars
situated at $(V-I)\sim 1.1$ and $V \sim 20$  mag might be HeB stars belonging to
the association or a residual population from the general field of the galaxy.
 In order to obtain the age of this association, we have fitted the position of the turnoff using a synthetic CMD  computed for this purpose. We find an age of $\sim$ 100 Myr for $Z=0.001$ and $\sim$ 125 Myr for  $Z=0.003$. Both synthetic diagrams are plotted over the observed CMD in Figure \ref{ob} (solid circles). Ortolani \& Gratton (1988) estimated an age of $\sim$ 10$^8$ yr for this association using its CMD---in good agreement with our result.

Finally, going back to the general field of Phoenix, a few bright blue stars are observed above the BP in the CMD of the inner region (Figure \ref{3cmd}a), with  20.0 mag $<I<$ 21.5 mag and $0 <(V-I) <0.5$. They
could be foreground stars, but the fact that a similar population is absent in the
CMD of the outer region (Figure 8c), where the foreground contamination should be larger, indicates they likely belong to Phoenix. Their position in the CMD is compatible with the bluest extent of the HeB phase of metal-poor intermediate-mass stars with ages in the range 150--500 Myr. This is shown in Figure \ref{vcmd}, where the position of the bluest extension of the blue-loop for different ages and metallicity $ Z=0.0004$ (filled points) and $Z=0.001$
 (open points) based on the Padua models are overplotted. This qualitatively indicates that the metallicity must be low. But no definitive conclusion can be drawn from this information alone because the maximum blue extension of HeB phase is very sensitive to both envelope  and core overshooting (Alongi {\it et al.} 1991).

\subsection {Spatial Variations of the Recent Star Formation} \label{recent}

The distribution of the BP and HeB populations can provide insights into the spatial variation of the recent star formation within the central component of Phoenix. To analyze this issue, we created stellar density maps for BP and HeB stars  following a method similar to that used by Alfaro {\it et al.} (1992) and Dohm-Palmer {\it et al.} (1997). The maps with the position of selected BP and HeB stars (see Sec. \ref{bp} and Figure \ref{mapa}) were convolved  with a two-dimensional Cauchy kernel, using a smoothing parameter $h_{x}=h_{y}=15$ pc. 
This density estimator is a way to discern actual features within
the data sample as well as a convenient way of presenting them. The errors in this procedure are discussed by  Alfaro {\it et al.} (1992).

The resulting stellar density maps for BP and HeB stars are plotted in Figure \ref{isomapa}. They show clear asymmetries in the distributions of the BP stars, which are predominantly located in the western half of the central component
of Phoenix, and
HeB stars, which are primarily situated in the eastern half. The offset of the centroid of the youngest population was first noted by Ortolani \& Gratton (1988) and  is also observed in the stellar map plotted in Figure~\ref{mapa}. 
Moreover, most of  BP stars are clustered in two associations clearly visible at $(X,Y)$ $\sim$ (700,680) and (620,720) in Figure \ref{isomapa}, indicating
that recent star formation has taken place  in localized regions of the galaxy. The maximum stellar density coincides with the association discussed in Sec. \ref{bp}, which is probably the youngest feature in the galaxy. This stellar distribution resembles those observed in dIrr galaxies ({\it e.g.,} Sextans A) rather than dEs and indicates that Phoenix may have looked like a typical dIrr some 100 Myr in the past.

 The asymmetry in the distribution of BP and HeB stars described above is also seen in the CMDs of the eastern half (Figure \ref{cmdisk}a) and West half (Figure \ref{cmdisk} b) of the central component. A more populated BP is observed in the western part of the galaxy, while HeB stars are predominantly situated in the eastern part.  This tendency is confirmed by the ratio of the number of  HeB to BP stars ($N_{\rm HeB}$, $N_{\rm BP}$) in both parts of the central component, which gives $N_{\rm HeB}/ N_{\rm BP}=1.1 \pm 0.28$ in the eastern half and $N_{\rm HeB}/ N_{\rm BP}=0.40 \pm 0.12$ in the western half. A possible explanation for this asymmetry in the central region of Phoenix is that interstellar dust might be obscuring stars in the eastern part of this region. However,  we have not found any signatures of dust in the CMDs of the inner region plotted in Figure \ref{cmdisk}, such as a radial color gradient among RGB stars or variation in the magnitude of the TRGB. Furthermore, there is no significant dust emission in the {\it IRAS} images near the position of Phoenix. For these reasons, it seems that the asymmetry would actually  be due to a difference in the recent SFR across the central component of the galaxy.

The observed gradients in the distribution of young stars suggest the global centroid of recent star formation has moved from east to west across the central component of Phoenix. This is consistent with a possible self-propagation of star formation. Interestingly, the cloud of H~{\sc i} detected by Young \& Lo (1997) and Oosterloo {\it et al.} (1998)  6$\arcmin$ south west from the center of the galaxy (see Sec. \ref{introduccion}) is close to the most recent star formation regions and might be involved in the process. If it belongs to Phoenix, it could provide good evidence that a burst of star formation can blow away the gas from a dwarf galaxy, preventing further star formation (Dekel \& Silk 1986). Assuming that the last burst of star formation is responsible of this H~{\sc i} cloud,  its expansion velocity can be calculated. Taking the projected distance between the cloud and the galaxy (720 pc), and the time of the last burst of star formation (100 Myr; see Sec. \ref{bp}), this gives an expansion velocity of $\sim$ 7 km s$^{-1}$. This value is similar to those observed in slowly expanding shells that dominate the ISM in some distant smaller galaxies, such as Holmberg I and M81 A dE (Puche \& Westpfahl 1994) and Holmberg II (Puche {\it et al.} 1992). In addition, the direction of the possible propagation of the star formation
in the central component of Phoenix is the same as that expected for the expansion of
the cloud.

If the gas cloud is connected with the star formation burst, the
question arises of whether the second could have blown the first away from
the central region of Phoenix. Considering the mass of the cloud
($1.2\times 10^5$ M$_\odot$; Young, \& Lo 1997) and the expansion
velocity deduced above, the kinetic energy involved in the expansion
is $6\times 10^{49}$ erg. This is a small portion of the total energy 
injected into the interstellar medium by a 10 $M_{\sun}$ star through stellar
winds or SN explosions ($10^{51}$ erg each; Leitherer 1998). On the other
hand, from the total mass converted into stars in the last Gyr in the inner
component of Phoenix ($10^{5 } M_{\sun}$; see Sec.\ref{sfh}) and using the IMF by Kroupa, Tout \& Gilmore (1993), about 300
stars more massive than 10  $M_{\sun}$ are expected have been formed, 
easily providing
 enough energy for the process.

The former reasoning is speculative, but it shows that it is extremely important to obtain the radial velocity of Phoenix ({\it e.g.,} through the
spectroscopy of its stars) in order to determine any possible association between
 the H~{\sc i} cloud and the galaxy.  H~{\sc i} has been also observed  in the external regions of Tucana dE (Oosterloo {\it et al.} 1996) and in Sculptor dE (Carignan {\it et al.} 1998), and could be present in other dEs for which past H~{\sc i} observations were centered on the optical image (Mateo 1998).

\subsection{ The intermediate-age population} \label{intermediate}

An open question about the SFH of galaxies which, like Phoenix, possess a predominantly old population but also show traces of recent star formation is whether the star formation has proceeded more or less continuously or, on the contrary, whether the young stars are the result of a singular event. There are two features in the CMD of Phoenix (Figure \ref{cmphoenix}) which are signatures of an  intermediate-age population: the upper-AGB stars and the RC of core helium-burning stars. We will use these to discuss the distribution of intermediate-age stars across the galaxy.

\subsubsection { The AGB population} \label{agb}

 The first evidence for the existence of intermediate age stars in Phoenix came from the observation by VDK of a significant number of stars brighter than the TRGB, which might be intermediate-age upper-AGB stars. Nevertheless, the lack of a CMD for a nearby comparison field prevented them from estimating the actual number of AGB candidates. This population is observed in our CMD (Figure \ref{cmphoenix}) extending up to $(V-I)\sim 3.0$, but the CMD of a comparison field situated  7.5$\arcmin$ east of Phoenix shows that the contamination by foreground stars is very high in this region of the CMD.

 In order to estimate the actual number of AGB stars as a function of the galactocentric radius, we have obtained the number of stars in two regions situated in the AGB locus of the CMD: i) 18 mag $<I<$ 19 mag, $1.4<(V-I)\leq 1.8$, which would contain the low-metallicity AGB stars and ii) 18 mag $<I<$ 19 mag, $(V-I)>1.8$, which would be populated by higher-metallicity AGB stars. The stellar ages would be intermediate and old in both cases. We will denote these numbers by $N_{\rm AGB,l}$ and $N_{\rm AGB,h}$ respectively. The results for the three CMDs plotted in Figure \ref{3cmd} and for the comparison field ($N_{f,l}$ and $N_{f,h}$) after scaling to the area of the corresponding region, are given in Table 3. The errors in the star counts are given according to Poissonian statistics. The results given in Table 3 show the stars above the TRGB can be easily accounted  for by the contribution of foreground stars. 

Nevertheless, there is evidence that some of these stars are actually  Phoenix members. Da Costa (1994) obtained low-resolution spectra of some  upper-AGB candidates, and found that two of them are carbon stars and thus actually belong to Phoenix. These stars are plotted in Figure \ref{mapa} as open circles and their magnitudes are $I_{0}=19.05$ mag and $I_{0}=18.84$ mag, respectively. Using the bolometric corrections by Da Costa \& Armandroff (1990) and a distance modulus of 23.0 (see Sec. \ref{distance}) we obtain $M_{\rm bol}= -3.51$ mag and $M_{\rm bol}=-3.69$ mag, which is in good agreement with Da Costa (1994). From a comparison of the magnitudes of these stars with those of carbon stars in the LMC and dE galaxies with substantial intermediate-age population, Da Costa (1994) concluded that the carbon stars in Phoenix are not likely to have ages much less than 8--10 Gyr. This value must be considered as only an
estimate
 because there may be bolometrically brighter (and thus younger) AGB stars that have not yet been identified.

\subsubsection {Gradient in the RC morphology} \label{redclump}

The RC of helium-burning stars is a stronger indicator of the presence of an intermediate-age population in the galaxy. Interestingly, a gradient is found in the number of stars populating it (Figure \ref{3cmd}).  Although this gradient is affected by crowding and other observational effects, we will show that it is produced by real differences in the RC population.

 To better illustrate this variation, it is useful to compare the observed CMDs with those recovered for the same regions in our analysis of the observational effects in Sec. \ref{crowding}. As the input synthetic CMD was computed assuming a constant SFR from 15 Gyr ago to date, any gradient found in the relative population of the RC of the recovered model CMDs can be interpreted as produced by observational effects. The recovered model CMDs for each region are shown in Figure \ref{cmdmodel}, in which a number of stars similar to that observed in the corresponding panel of Figure \ref{3cmd} has been plotted. It can be seen that the RC is prominent at all three galactocentric distances, with the crowding actually making the RC less populated relatively to the whole CMD in the central region (its total number of stars is practically the same as in the intermediate region).   

To quantify this gradient, we have calculated the observed ratio of RGB to RC stars across the galaxy and the corresponding values for the model CMDs. We have defined two regions in the CMDs: i) a RC region (23 mag $ < I < $ 22 mag,
 $0.8 < (V-I) < 1.2$) and ii) an RGB region (21 mag $ < I <$ 22 mag, $1.0 < (V-I) < 1.3$). We will denote the number of stars in them as $N_{\rm RC}$ and $N_{\rm RGB}$ respectively. The  $N_{\rm RGB}/N_{\rm RC}$ ratio can be used to test the gradient in the RC morphology with galactocentric radius. However, this observed ratio is very sensitive to the observational effects which mainly
affect it  in two ways (see Aparicio \& Gallart 1995): (1) the loss of stars is more important at fainter magnitudes, and thus the number of RC stars is underestimated in the inner regions of the galaxy, raising the $N_{\rm RGB}/N_{\rm RC}$ ratios; and (2) the migration of RC stars to the RGB region  produced by observational effects also contributes to an increase in the $N_{\rm RGB}/N_{\rm RC}$ ratio. In order to understand the role of crowding and other observational effects on the variation of the RC morphology along the galaxy, we compared the $N_{\rm RGB}/N_{\rm RC}$ ratios of the observed (Figure \ref{3cmd}) and model (Figure \ref{cmdmodel}) CMDs. Table 4 lists the results: column 1 gives the region ; columns 2 to 4  list the results for the observed CMDs, and column 5 shows the ratios of RGB to RC stars for the model CMDs ($N_{\rm RGB}^{\rm s}/N_{\rm RC}^{\rm s}$). 

 The synthetic $N_{\rm RGB}^{\rm s}/N_{\rm RC}^{\rm s}$ decreases with galactocentric radius, according to what is expected from the fact that observational effects
are more severe in  region A. However, the observed $N_{\rm RGB}/N_{\rm RC}$ ratio increase with the galactocentric radius.  This gradient is compatible with two different
scenarios: a) the presence of a substantial intermediate-age population
in the central region of Phoenix, which would be less prominent in the
outer regions; and/or b) variation in the metallicity
of the old population with galactocentric radius, in which the more
metal-rich old stars are concentrated in the inner regions of the
galaxy, hence moving helium-burning stars from the HB into the RC. However, a gradient in the metal abundance of the old population of Phoenix would produce a change in the position, inclination and/or width of the RGB which is not observed in our CMDs (see Sec. \ref{metallicity}). Therefore, the variation of the  $N_{\rm RGB}/N_{\rm RC}$ ratio supports the existence of a larger intermediate-age population in the inner regions of the galaxy, and thus a more complex and extended SFH than in the  outer regions.

Details of short-lived star formation bursts cannot be given from the previous
synthetic CMDs analysis, but the similarity of the values of  $N_{\rm RGB}/N_{\rm RC}$ and $N_{\rm RGB}^{\rm s}/N_{\rm RC}^{\rm s}$ for region A suggests that, on average, the general trend of the star formation scenario for this region is not far from the constant SFR assumed for the model. Instead, the differences found for regions B and C indicate that this is not the case for the outer part of the galaxy, where an SFR weighted towards older ages is more likely.
 
\subsection{The old population} \label{old}
 
\subsubsection{Gradient in the HB morphology} \label{hb}

The CMDs of the different regions of Phoenix plotted in Figure \ref{3cmd} display variation in the HB morphology from the center to the external parts of the galaxy. In the inner region (Figure \ref{3cmd}a), the HB is very poorly defined and its blue part merges with the {\it blue plume} of young stars, making it difficult to establish its actual extension. However, the HB appears well-defined and extends up to $(V-I)\sim 0$ in the outer, less crowded region of Phoenix (Figure \ref{3cmd}c). The presence of such blue
HB stars is a strong indication that an old, low-metallicity population, of an age comparable to those of the oldest Galactic globular cluster, exists in the outer part of Phoenix.

Unfortunately, the HB is very close to the limit of our photometry and it is affected by large observational errors. In fact, at these faint magnitudes, the changing crowding conditions may produce significant variations in the appearance of the HB as a function of galactocentric radius. This effect is  illustrated in Figure \ref{cmdmodel}. These diagrams suggest the gradient in the HB morphology observed in Figure \ref{3cmd} could, at least partially, be an artifact of the  crowding gradient across the galaxy rather than a real variation of the age or metallicity of the oldest stellar population. 

To   analyze this issue quantitatively,  we obtained the ratio of the HB to RGB stars for these three CMDs, following the same method used to study the RC morphology in Section \ref{redclump}. In this case, star counts and crowding factors in CMD regions situated in the HB (24 mag $< I <$  23 mag, $0.2 < (V-I) < 0.6$) and RGB (same as in Sec. \ref{redclump}) were computed to derive the $N_{\rm HB}/N_{\rm RGB}$ ratio in each CMD and the corresponding $N_{\rm HB}^{\rm s}/N_{\rm RGB}^{\rm s}$ (Figure \ref{cmdmodel}). The results  are listed in Table 5 where column 1 gives the region, columns 2  to 4 refer to the observed CMDs, and column 5 gives the HB to RGB ratio for the model CMDs corresponding to
each region. The last shows that the $N_{\rm HB}^{\rm s}/N_{\rm RGB}^{\rm s}$ ratio decreases towards the inner part of the galaxy, as a result of the 
observational effects in the center of the galaxy.  The relatively small variations in the $N_{\rm HB}^{\rm s}/N_{\rm RGB}^{\rm s}$ ratio in the regions B and C indicate that the large differences in the same ratio observed  in the galaxy for the same regions (column 4) cannot be explained as observational effects. These differences are by themselves enough indication that traces of
a HB are being detected. But not only that. Although we are at the limit of the photometry and consequently any conclusion must be taken with care, the diferences in the HB relative strength might indicate a gradient in their properties. Leaving aside region A, for which the HB region is probably contaminated by BP stars as well, we can speculate a little bit more: if the
apparent increasing of relative strength of the HB for larger radii is real, it 
could indicate the existence of an older, metal-poor spheroid surrounding the central component of the galaxy. Although this cannot be considered as a definitive conclusion, we propose it as a future test to be done in Phoenix.

\subsubsection{Radial distribution of the metallicity and age} \label{metallicity}

For stars older than 2 Gyr, the effect of age on the position of the RGB is much smaller than the effect of metallicity. For this reason, if the age dispersion is not too large, the metallicity dispersion can be estimated from the wideness of the RGB (Da Costa \& Armandroff 1990). However this might not be the case if stars with a large interval of ages are present. A clear example of this is Leo~I, a galaxy
similar to Phoenix in which all the dispersion of the RGB can be accounted for by the age dispersion alone  because the metallicity seems to be nearly constant for the entire life of the galaxy (Gallart {\it et al.} 1999b). This could also be true for Phoenix. However, we will explore the position and dispersion of the RGB, looking for possible gradients without making any {\it a priori} assumptions as to whether they are produced
by metallicity or age differences.

We have estimated the median $(V-I)_{-3.5,0}$ and the color dispersion, $\sigma (V-I)_{-3.5}$, from the RGB stars at different radii. With this purpose, RGB stars  with 19.25 mag $<I<19.75$ mag were selected in each region, carefully
 removing  the foreground stars using statistical field-star subtraction by means
of the comparison
field. The color dispersion was computed as the standard deviation,  $\sigma(V-I)_{-3.5}=[{\rm med}(|\delta_{j}^{c}-{\rm med}(\delta_{j}^{c})|]/0.675$, where $med$ stands for the median and $\delta_{j}^{c}$ is the difference between the color of the star $j$  and the color of the straight line that best fits the RGB evaluated at magnitude of the star $j$. This estimate provides a more accurate way of measuring  the width of the RGB, because it takes into account the effect of the slope of the RGB, which  would otherwise tend to increase the color dispersion.

 The results are summarized in Table 6: column 1 gives the region of the galaxy, columns 2 and 3 give the median color and the color dispersion of the RGB stars, respectively, and columns 4 and 5 give the metallicity and metallicity dispersion derived from the $(V-I)_{-3.5,0}$ and  $\sigma (V-I)_{-3.5}$, respectively. There are no significant variations in either the mean color of the RGB stars or its width for the three regions, indicating that, in contrast to what has been found for the younger stars, the population older than 2 Gyr is uniformly distributed across Phoenix, or at least there is no evidence contrary to this hypothesis. For illustrative purposes only, the metallicity, $[{\rm Fe/H}]$, and the metallicity 
dispersion have been computed as if the color dispersion of the RGB were due
to metallicity effect only. For this purpose, the calibration of Da Costa \& Armandroff (1990) has been used. The results for [Fe/H] and $\sigma([{\rm Fe/H}])$ are listed in columns 4 and 5 of Table 6. In summary,  the
mean metallicity is [Fe/H] $=-1.37$, that is significantly larger than that
obtained by VDK ([Fe/H] $=-2.0$). To further test this issue we have compared directly our RGB with the RGB loci of globular clusters by Da Costa \& Armandroff (1990). The result is shown inf Fig. \ref{cmgb}. It can be seen in this figure that the slope of our RGB seems to be in better agreement with those of slightly more metal poor clusters ($[Fe/H]\sim -1.6$ or lower). If the slope of the RGB is related with the metallicity only, some alternative pernicious effect should be
at work. There are two possible explanations for this apparent disagreement. Firstly, a zero-point systematic error should be present in our $V$ photometry (producing a  shift in $(V-I)$). Although this is possible, we believe that
is it unlikely because it should be as big as $\sim 0.15$ magnitudes, which is really too much considering the cautious photometric calibration we have done, based
on a large number of standard stars measures (see Sec. \ref{photo}). Alternatively, the disagreement with the globular cluster fiducial RGB might
originate in a wrong reddening. In such a case, the reddening should be $E(V-I)\sim 0.15$ or $A_{I}\sim 0.3$, which would result in a shorter distance modulus of $(m-M)_{o}\sim 22.77$ or $d\sim 360$ kpc, but that would leave mostly unchanged the remaining conclusions of the paper.

\subsection {Clues on the Star Formation History} \label{sfh}

	We have estimated the average SFR for the last 1 Gyr ($\bar{\psi}_{< 1 {\rm Gyr}}$) and for the age interval 1--15 Gyr ($\bar{\psi}_{>1 {\rm Gyr}}$) from the amount of BP and red tangle stars observed in the CMDs, in a way similar to Aparicio {\it et al.} (1997c). For this purpose, we have obtained the star counts in the BP (as representative of stars younger than 1 Gyr) and in the red tangle (which contains stars of all ages in the interval $\sim$1--15 Gyr). The SFR values were estimated by scaling the relative number of blue and red stars to those observed in a model CMD computed with constant SFR from 15 to 0.1 Gyr and a linear chemical enrichment law from $0.001<Z(t)<0.003$. This metallicity range adequately reproduces  the position and width of the RGB for this age interval.

 The results are summarized in Table 7, which gives the number of red and blue stars, the ratio between them, the derived $\bar{\psi}_{>1 {\rm Gyr}}$ and $\bar{\psi}_{< 1 {\rm Gyr}}$ (in $M_{\sun}$ yr$^{-1}$) for each region of the galaxy and the same after normalization to the area of each region in pc$^{2}$, $\bar{\psi}_{>1 {\rm Gyr}}$/A and $\bar{\psi}_{< 1 {\rm Gyr}}$/A (in $M_{\sun}$ yr$^{-1}$ pc$^{-2}$). The most interesting result is that Phoenix has been  forming stars in its central region during the last Gyr at the same average SFR rate as the average SFR over its entire lifetime, while it has dramatically decreased in the last Gyr for the two outer regions defined in the galaxy.

As discussed in Sec. \ref{recent}, the distribution of BP stars suggests that $\bar{\psi}_{< 1 {\rm Gyr}}$/A  varies across the central component of the galaxy. The most active region is the western one, in which the young association lies (see Sec.\ref{bp}).The total mass involved in the burst that produced this association can be also calculated from our models, resulting to be $1.8 \times
10^{4}\ M_{\sun}$. This is 20 \% of the total mass converted into stars in the last 1 Gyr in the inner component of Phoenix, and corresponds to 2.1 $M_{\sun}$ pc$ ^{-2}$ for the present-day area covered by the association ($8.5 \times 10^{3}$ pc$^{2}$).

In the outer regions, the recent SFR ($\bar{\psi}_{< 1 {\rm Gyr}}$ and $\bar{\psi}_{< 1 {\rm Gyr}}$/A) is much lower than the average past SFR ($\bar{\psi}_{>1 {\rm Gyr}}$ and $\bar{\psi}_{>1 {\rm Gyr}}$/A). In fact, the recent SFR values of these regions are only upper limits because of the contamination of foreground stars\footnote{ Unfortunately, our comparison field CMD is not deep enough to decontaminate the star counts in the
 BP region.}. Perhaps star formation is not taking place at all in these regions. It is also interesting to note that in the outermost region, the average SFR is lower. This is consistent with the gradient in the RC morphology found in Section \ref{redclump}, which would be compatible with the star formation stopping at earlier times in the outer regions of the galaxy.

\section{Variable stars} \label{variable}
 
The low metallicity of Phoenix and the age structure derived from its CMD make it a particularly interesting galaxy for studying the characteristics of 
short-period, low-metallicity Cepheid variables, which can provide additional insight into the period-luminosity (PL) relation at the low-mass, low-metallicity extreme. On the other hand, Phoenix is also a good candidate for the discovery of anomalous Cepheids (ACs), which have been only found in the dE galaxies and in the SMC. Both types of variable stars seem to show important similarities in their low-mass, low-metallicity extreme (Smith {\it et al.} 1992).

No systematic variable star surveys have been carried out in Phoenix. Caldwell \& Schommer (1988) reported the start of a Cepheid survey in Phoenix using the Cerro Tololo 4-m telescope, but the details of their work have not yet been published. They found several variable stars, and derived an apparent distance modulus of 24.7.  VDK identified a small number of variable candidates by comparing their magnitudes in  frames taken on two different nights. Only one of the stars reported by VDK (number 157 in their Table 3) is clearly confirmed as variable in our data in the $V$ band, while the rest of candidates did not exhibit significant magnitude variations during our two consecutive nights. 

We searched our data for variables by first selecting the stars close to the Cepheid instability strip given by Chiosi, Wood \& Capitanio (1993).  The magnitudes of these candidates were transformed into a common instrumental system using DAOMASTER (Stetson 1997). This list was then used to search for variable stars, selecting only those stars which displayed luminosity variations larger than 0.3 mag. The list of variable candidates is given in Table 8: column 1 gives the number of the stars in our photometry list, columns 2 and 3 give the $x$ and $y$ coordinates in pixels, columns 4 and 5 list the mean $V$ and  $I$ magnitudes, column 6 lists the mean $(V-I)$ color index, column 7 gives the mean absolute magnitude in the $I$ band (which in random phase observations represents the mean magnitude of a Cepheid better than $V$; Freedman 1988),
and column 8 gives the observed $V$ variation in magnitudes and its error,

The position of these candidates in the CMD of Phoenix is shown in Figure \ref{cmcefeida}, together with the locus of the Cepheid instability strip. The location is compatible with those of low-luminosity classical Cepheids, although they are rather faint. For this reason, it is more likely they are AC or W Vir stars. Light curves are necessary to obtain a reliable classification.

Figure \ref{chart} displays the field of these variable candidates, showing their brightness variation.  Interestingly, the position of the four variable stars is very close to the eastern border of the central component of Phoenix (see also Figure \ref{mapa}). This suggests that these short-period variable candidates could be tracing a stellar population of a particular age. In fact, if they were classical Cepheids, the age gradient across the central component of Phoenix found in Sec.\ref{recent} would produce a gradient in the number and the periods of these variables. In this sense, a larger number of short-period variables  would be found predominantly in the eastern part of the central component, which contains the older HeB stars (see Figure \ref{isomapa}), while longer-period, younger variables would be placed in the western part. In fact,  their
longer periods would make it more difficult to detect them in our two-day
search. A more complete survey and the knowledge of the nature of these variables (whether classical or anomalous Cepheids) would help to achieve a definitive answer.

In summary, we identify several short-period variable star candidates in Phoenix which could be AC or W Vir. Since we have  compared data only for two consecutive nights, we are probably missing many candidates, although the number of BP stars in our CMD suggests that classical Cepheids cannot be very abundant.  We are presently carrying out a systematic survey to characterize the variable star population of this galaxy. The results will be presented elsewhere (Gallart {\it et al.} 1999c).

\section{ Globular clusters} \label{globular}

Canterna \& Flowers (1977) identified three globular
cluster (GC) candidates 1.5$\arcmin$ south east of the center of Phoenix, on the basis of their non-stellar appearance. Our best quality images show that objects $\#$1 and $\#$2 (see their figure 2) are clearly background galaxies. However, the classification of object $\#$3 is more doubtful, due to its more compact structure and quasi-circular morphology. In addition, we identified three other similar (compact, circular but nonstellar) objects near the center of Phoenix. The finding chart for these GC candidates is displayed in Figure \ref{globulares} (marked  a, b and c), together with the objects selected
by Canterna \& Flowers ($\#$1, $\#$2 and $\#$3).

Integrated $V$ and $I$ magnitudes were obtained for the three objects identified by us and the $\#$3 object identified by Canterna \& Flowers (1977) after subtracting the neighboring stars from the frame. Table 9 lists the results: column 1 identifies the cluster, columns 2 and 3 list the $x$ and $y$ coordinates in pixels, columns 4 and 5 give the $V$ magnitude and the $(V-I)$ color index, column 6 gives the absolute magnitude $\langle M_{V}\rangle $, and column 7 gives the FWHM. The mean $V$ magnitude of these four objects is $\langle V\rangle =18.85$ mag (with $\sigma=0.15$), which corresponds to $\langle M_{V}\rangle =-4.20$ mag, and the mean color is $\langle (V-I)_{o}\rangle =1.43$. Local Group GCs display a luminosity distribution with $\langle M_{V}\rangle =-7.1$ mag, $\sigma \sim 1.3$ (independent of the luminosity of the parent galaxy or other factors such as galaxy type or environment; Harris 1991), and  $0.7<(V-I)_{o}<1.2$. The GC candidates in Phoenix are therefore redder and about three magnitudes fainter than the average in the Local Group.  This could indicate that they are background galaxies, in spite of their compact structure.

The definitive answer about the nature of these objects will come soon with the deep {\it HST} data scheduled for the central part of Phoenix. If they are confirmed to be background galaxies, the lack of GCs in Phoenix would be in agreement with observational result that suggests there is a minimum luminosity for a dwarf galaxy to form globular clusters. The lowest luminosity galaxies in the Local Group containing GCs are Fornax ($N_{\rm GC}$=5) and Sagittarius ($N_{\rm GC}$=4), with $M_{V}=-13.2$ mag  and $M_{V}=-13.4$ mag, respectively, (Mateo 1998) while Phoenix ($M_{V}=-10.1$ mag) is much fainter.

\section{Conclusions}  \label{conclusion}

We present new deep $VI$ ground-based photometry of the Local Group dwarf galaxy Phoenix. Our results show that this galaxy is mainly dominated by red stars, with some blue, MS, and BL stars indicating recent (100 Myr old) star formation in the central part of the galaxy. The large area covered in our study allows us to investigate possible gradients in its stellar populations. We have performed an analysis of its structural parameters based on a scanned ESO/SRC plate, in order to search for possibly differentiated components in the galaxy. The elliptical isopleths  show a sharp rotation of $\simeq 90 \deg$ in their major axes at radius $r\simeq 115\arcsec$ ($\sim$ 230 pc) from the center, suggesting the existence of two components: an inner one oriented   east--west, which contains all the young stars of the galaxy, and an outer one oriented   north--south, which seems to be  predominantly populated by old stars. These results have then been  used to obtain CMDs for three different regions in order to study the gradients in the stellar population properties within the galaxy.

The young population located in the central component of Phoenix shows a clear asymmetry in its distribution, with the younger BP stars predominantly located in the western half of the central component and the relatively older HeB stars predominantly situated to the east. This spatial variation could indicate a propagation of star formation across the central component. The H~{\sc i} cloud found  $\sim6\arcmin$ south west by Young \& Lo (1997) could have been involved in this process, and might be good evidence that a burst of star formation can blow away the gas from a dwarf galaxy, preventing further star formation. However, the optical radial velocity of Phoenix is necessary to confirm the association of this cloud with it. We also found a gradient in the intermediate-age population based on the ratio between the number of stars in an area of the RGB, $N_{\rm RGB}$, and an area of the RC, $N_{\rm RC}$, in the CMD. Because no metallicity gradient is observed in the old population of Phoenix, this gradient indicates the presence of a substantial
intermediate-age population in its central region that would be
less abundant or absent in the outer regions. This result is also consistent  with the gradient found in the ratio between $N_{\rm RGB}$ and the number of stars in an area of the HB, $N_{\rm HB}$. This ratio indicates that
the relative number of HB stars increases towards the outer part of the galaxy, and that therefore this outer region  might be older on average. 

The former result suggests the existence of an older, metal-poor population distributed in a spheroid surrounding the central component of Phoenix, resembling the halo--disk structure observed in the Local Group spirals. We propose three main possibilities  for the nature of this population, which would be indistinguishable with the current data: i) all the stars in the outer part were formed in a short initial period, which would produce a predominantly old stellar population, like in the Milky Way; ii) a time-extended SFH has taken place all over the galaxy, but stopped  several Gyr
ago in the external area, which result in no young stars, but a mixture
of old and intermediate-age stars in the outer part; or iii) the SFR has always been low at any time in the outer regions, resulting in very few or no young stars there.

It would be interesting to study other properties of the outer  versus the inner part of the galaxy, like the metallicity and abundance ratios as a function of radius and the kinematic properties. This would allow a determination of whether a halo-like population (older and kinematically differentiated) exists in Phoenix. In addition, {\it HST} data of the outer regions would provide excellent information on its age structure and be crucial for the characterization of the possible halo population surrounding Phoenix.

We have estimated the average  SFR for the last 1 Gyr ($\bar{\psi}_{< 1 {\rm Gyr}}$) and for the age interval 1-15 Gyr ($\bar{\psi}_{>1 {\rm Gyr}}$) from the amount of the blue and red stars observed in the CMD. Our results indicate that  overall Phoenix evolved with an approximately constant SFR on average in its central region, in agreement to that observed in most dIrr galaxies (Mateo 1998). This behavior is quite similar to the Pegasus dIrr, which also has an average SFR very similar to Phoenix ($\bar{\psi}$/A = 1.5 $\times$ 10$^{-9}$ $M_{\sun}$ yr$^{-1}$ pc$^{-2}$, Aparicio {\it et al.} 1997a). This is not the case of LGS 3, considered along with Phoenix as a prototype of the intermediate case between dIrr and dE galaxies in the Local Group. In this galaxy, the recent SFR is three times less than its lifetime average (Aparicio {\it et al.} 1997b).

The recent SFR of Phoenix is also comparable to those of typical dIrr galaxies except perhaps for the fact that Phoenix lacks any strong very recent bursts like those exhibited by galaxies such as Sextans A (Aparicio {\it et al.} 1987; Dohn-Palmer {\it et al.} 1997) or NGC 6822 (Gallart {\it et al.} 1996b, c). We have also compared our result for the recent SFR of Phoenix with the SFR values (normalized to the total area) for dIrr galaxies compiled by Hunter \& Gallagher (1986). These values were determined from the $H_{\alpha}$ luminosity of the galaxy, and are therefore a measure of the current SFR. The normalized SFR for the central region of Phoenix ($1\times 10^{-9} M_{\sun}$ yr$^{-1}$ pc$^{-2}$) is in the range obtained for this sample ($2.5 \times 10^{-11}\ M_{\sun}$ yr$^{-1}$ pc$^{-2}  <  \psi/A < 1.6 \times 10^{-9}\ M_{\sun}$ yr$^{-1}$ pc$^{-2}$).

We determined a distance modulus for Phoenix of $(m-M)_0=23.0\pm 0.1$, using the tip of the red-giant branch (TRGB) as a distance indicator. Finally, we found four short-period variable candidates from our photometry, that might be anomalous Cepheids or W Vir stars. Moreover, it is very unlikely  Phoenix has globular clusters, as  is expected for a galaxy with this faint absolute magnitude.

\acknowledgments

We are grateful to M. Irwin for scanning a photographic plate of Phoenix for
our  study of its structural parameters. We thank to the anonymous referee for careful reading of the original manuscript and for useful comments and suggestions. We also thank to E. J. Alfaro and J. Cabrera-Ca\~no for their help with the kernel method used in this paper, and L. Verdes-Montenegro for allowing us to use her algorithm for fitting ellipses. 
 This work has also been substantially improved by fruitful discussions with G. Bertelli, G. Da Costa, S. Demers,  C. Esteban, W.L. Freedman, M. Mateo, D. I. M\'endez, T. Oosterloo, M. Trewhella, and L. Young. This work is part of the PhD thesis
of DMD, and has been financially supported by the Instituto de Astrof\'\i sica de Canarias (grant P3/94) and the Direcci\'on General de Ense\~nanza Superior, Investigaci\'on y Desarrollo of the Kingdom of Spain (Grant PB94-0433).

\newpage

\figcaption[fig1.ps]{Color image of the central region of Phoenix, obtained
as a combination of $B$, $V$ and $I$ images. \label{color}}

\figcaption[fig2.ps]{Standard error of the mean, as a function of the magnitudes provided by ALLFRAME for the stars included in the final photometric list. \label{error}}
 
\figcaption[fig3.ps]{Injected and recovered magnitudes for the artificial stars from our analysis of the observational effects in Phoenix. The stars in a synthetic CMD constructed with constant SFR (from 15 to 0.01 Gyr) and $Z$ from 0 to $Z=0.002$ have been used as the artificial stars.\label{artificial}}

\figcaption[fig4.ps]{Center portion of the UKST plate used in
the analysis of the spatial structure of Phoenix. The exposure time was 60 min and the limiting magnitude is $B\sim 22$. The field is 35$\arcmin$ $\times$ 35$\arcmin$. North is at the top and east to the left. \label{plate2}}

\figcaption[fig5.ps]{Isopleth map of Phoenix obtained from star counts in an  ESO/SRC plate. The size of the region is $11.5\arcmin \times 10.8\arcmin$.
The first contour is 15 arcmin$^{-2}$ and the increment is 5 arcmin$^{-2}$. North is at the top and east is to the left. \label{isopleth_fig}}

\figcaption[fig6.ps]{Radial number density of stars in Phoenix after correcting for both crowding effects and foreground contamination.\label{king2}}

\figcaption[fig7.ps]{The square root of the stellar surface density is plotted
as a function of the inverse of the semimajor axis of the elliptical isopleths (dots).
The tidal radius of Phoenix, as defined by King (1962), can be obtained from the slope and zero-point of the linear least-squares fitting (solid line) . \label{king}}

\figcaption[fig8.ps]{[$(V-I)$, $I$] CMD for Phoenix. A total number
of 6530 stars are plotted.\label{cmphoenix}}

\figcaption[fig9.ps]{$I$ luminosity function for the red stars of Phoenix (solid line). The convolution of the luminosity function with a Sobel filter is represented by a dotted line. \label{tip}}

\figcaption[fig10.ps]{The limits for the three regions of Phoenix selected
in Sec. 6 to study the gradient in the stellar population are overplotted to a $V$ CCD image of the galaxy. The total field of the frame is $8\arcmin \times 8\arcmin$. North is at the top and east is to the left.\label{image1}}

\figcaption[fig11.ps]{CMDs for different regions of Phoenix: region A, 0 pc $< a_{d} \leq$ 229 pc; region B,  229 pc$ < a_{h} \leq$ 457 pc; region C, $a_{h}>$ 457 pc. \label{3cmd}}

\figcaption[fig12.ps]{Finding chart for the different young stellar populations selected from the CMD of Phoenix. The solid circles are BP stars,  triangles are HeB stars with ages $\sim$ 0.5 Gyr, and crosses are variable star candidates. The positions of two carbon stars confirmed by Da Costa (1994) are shown as  open circles. The boundaries for the proposed central and outer components of Phoenix and for the association are also displayed. North is at the top
 and east to the left. \label{mapa}}

\figcaption[fig13.ps]{[$(V-I)$, $V$] CMD of the central region of Phoenix (region A). An isochrone from the Padua library (Bertelli {\it et al.} 1994) of  $Z=0.001$ and age 800 Myr is overplotted. The bluest extent of the HeB phase for metal-poor stars with $Z=0.0004$ (filled circles) and $\log ({\rm age})= 8.2-8.6$, and $Z=0.001$ (open circles) and $\log ({\rm age})= 8.3-8.7$ (open circles) are also shown. The increment in $\log ({\rm age})$ is 0.1 in both cases. \label{vcmd}}

\figcaption[fig14.ps]{CMD of the young association situated on the western border of the central region (open circles). The solid circles correspond to the synthetic CMDs for metallicity $Z=0.001$ and age=100 Myr (left panel), and $Z=0.003$ and age=125 Myr (right panel).\label{ob}}

\figcaption[fig15.ps]{Stellar density maps for BP and HeB stars within the central component of Phoenix. The field is  260 pc$^2$. The first contour is 1.4$\times 10^{-3}$  pc$^{-2}$ and the increment 7 $\times 10^{-4}$ pc$^{-2}$. The ellipse marks the limits of the central component of the galaxy. \label{isomapa}}

\figcaption[fig16.ps]{CMDs for the eastern  and  western halves of the central component of Phoenix: a) Eastern part; b) western part. \label{cmdisk}}

\figcaption[fig17.ps]{CMDs with the magnitudes and colors of the artificial stars as recovered in our artificial-star tests, for the three regions of the galaxy as in Figure 9. The number of stars in each CMD is the same as the number of stars in the observed CMD for the corresponding region (Figure 9). \label{cmdmodel}}

\figcaption[fig18.ps]{Comparison of the observed RGB loci  of Phoenix with those of Galactic globular clusters by Da Costa \& Armandroff (1990). The solid curved lines are from the left to the right the position of the giant branches of
M15, M2, NGC 1851 and 47 Tuc, the metallicities of which are, respectively,
 [Fe/H] = --2.17, --1.58, --1.29, and --0.71 dex. \label{cmgb}}

\figcaption[fig19.ps]{ Position of the variable stars candidates in the CMD of Phoenix. The Cepheid instability strip is also shown for $(m-M)_{0}=23.0$.\label{cmcefeida}}

\figcaption[fig20.ps]{$V$ images of the central field of Phoenix taken on two consecutive nights: February 9 (left) and February 10 (right). A significant  change in the magnitude of the variable stars candidates listed in Table~8 is observed between both nights.\label{chart}}

\figcaption[fig21.ps]{Finding chart for the globular cluster candidates listed in Table 9.\label{globulares}}
\newpage

\begin{deluxetable}{cccccc}
\tablenum{1}
\tablewidth{0pt}
\tablecaption{ Journal of observations}
\tablehead{
\colhead{Date}      &
\colhead{UT}          & \colhead{Filter}  &
\colhead{Exp. time (s)}        
&\colhead{ FWHM ($\arcsec$)}}
\startdata
 9/02/97 & 01:34 & $I$ & 300 & 1.0 \nl
10/02/97 & 01:07 & $V$ & 300 & 1.0 \nl
10/02/97 & 01:14 & $V$ & 300 & 1.0 \nl
10/02/97 & 01:21 & $I$ & 300 & 0.9 \nl
10/02/97 & 01:27 & $I$ & 300 & 1.0 \nl
10/02/97 & 01:34 & $I$ & 300 & 0.9 \nl
10/02/97 & 01:40 & $V$ & 300 & 1.0 \nl
11/02/97 & 00:44 & $V$ & 300 & 1.0 \nl
11/02/97 & 00:50 & $V$ & 300 & 1.0 \nl
11/02/97 & 00:57 & $V$ & 300 & 0.8 \nl
11/02/97 & 01:10 & $I$ & 600 & 0.9 \nl
11/02/97 & 01:21 & $I$ & 600 & 0.9 \nl
11/02/97 & 01:33 & $I$ & 600 & 0.9 \nl
11/02/97 & 01:45 & $V$ & 300 & 1.0 \nl

\nl
\enddata
\end{deluxetable}
\newpage

\begin{deluxetable}{lcccr}
\tablenum{2}
\tablewidth{0pt}
\tablecaption{Structural parameters of Phoenix }
\tablehead{     
 \colhead{ } 
& \colhead{X$_{\rm cen}$}
& \colhead{Y$_{\rm cen}$}
& \colhead{$\epsilon$}& \colhead{PA($\arcdeg$)}}

\startdata
 Central component & 1052 & 998 & 0.4 & 95 \nl
 Outer component & 1051 & 999 & 0.3 & 5 \nl
\enddata
\end{deluxetable}

\begin{deluxetable}{ccccc}
\tablenum{3}
\tablewidth{0pt}
\tablecaption{AGB stars and contamination by foreground stars}
\tablehead{
\colhead{ } &    
\colhead{$N_{\rm AGB,h}$} & 
\colhead{$N_{\rm AGB,l}$} &
\colhead{$N_{f,h}$} & 
\colhead{$N_{f,l}$}}

\startdata
Region A & 4 $\pm$ 2 & 5 $\pm$ 2 & 6 $\pm$ 1 & 2 $\pm$ 0 \nl
Region B & 16 $\pm$ 4 & 6 $\pm$ 2 & 18 $\pm$ 2 & 6 $\pm$ 1 \nl
Region C & 23 $\pm$ 5 & 5 $\pm$ 2 & 39 $\pm$ 5& 14 $\pm$ 2 \nl
\enddata
\end{deluxetable}

\begin{deluxetable}{ccrcc}
\tablenum{4}
\tablewidth{0pt}
\tablecaption{ Gradient in the Red Clump }
\tablehead{
\colhead{ } &    
\colhead{$N_{\rm RGB}$} & 
\colhead{$N_{\rm RC}$} &  
\colhead{$N_{\rm RGB}/N_{\rm RC}$} & 
\colhead{$N_{\rm RGB}^{\rm s}/N_{\rm RC}^{\rm s}$}}

\startdata
Region A & 293 & 906  & 0.32 $\pm$ 0.03  & 0.35 $\pm$ 0.03 \nl
Region B & 420 & 1033  & 0.41 $\pm$ 0.03 & 0.25 $\pm$ 0.02\nl
Region C & 190 & 412 & 0.46 $\pm$ 0.06 & 0.22 $\pm$ 0.01\nl
\enddata
\end{deluxetable}
\newpage

\begin{deluxetable}{crccc}
\tablenum{5}
\tablewidth{0pt}
\tablecaption{ Gradient in the Horizontal Branch }
\tablehead{
\colhead{ } &    
\colhead{$N_{\rm HB}$} & 
\colhead{$N_{\rm RGB}$} & 
\colhead{$N_{\rm HB}/N_{\rm RGB}$} & 
\colhead{$N_{\rm HB}^{\rm s}/N_{\rm RGB}^{\rm s}$}}

\startdata
Region A & 116 &  293  & 0.40 $\pm$ 0.06 & 0.11 $\pm$ 0.03 \nl
Region B & 119 &  420  & 0.28$\pm$ 0.04 & 0.20 $\pm$ 0.03 \nl
Region C &  83 &  190 & 0.44 $\pm$ 0.08 & 0.18 $\pm$ 0.02\nl
\enddata
\end{deluxetable}

\begin{deluxetable}{ccccc}
\tablenum{6}
\tablewidth{0pt}
\tablecaption{ Radial variation of metallicity from RGB stars }
\tablehead{\colhead{} &    
 \colhead{$(V-I)_{-3.5,0}$}
 & \colhead{$\sigma (V-I)_{-3.5}$} 
& \colhead{$[{\rm Fe/H}]$} & \colhead{$\sigma([{\rm Fe/H}])$}}

\startdata
Region A & 1.46 & 0.06 & -1.33 & 0.34 \nl
Region B & 1.46 & 0.06 & -1.33 & 0.34 \nl
Region C & 1.44 & 0.06 & -1.39 & 0.36 \nl
\enddata
\end{deluxetable}

\newpage

\begin{deluxetable}{crrccccc}
\tablenum{7}
\tablewidth{0pt}
\tablecaption{Star formation rates }
\tablehead{\colhead{  }     
 & \colhead{$N_{\rm blue}$} &  \colhead{$N_{\rm red}$} 
 & \colhead{$N_{\rm blue}/N_{\rm red}$}
 & \colhead{$\bar{\psi}_{\rm >1Gyr}$}
 & \colhead{$\bar{\psi}_{\rm <1Gyr}$}
 & \colhead{$\bar{\psi}_{\rm >1Gyr}$/A}
& \colhead{$\bar{\psi}_{\rm <1Gyr}$/A}}

\startdata
Region A & 140 & 2026 & 0.070 & 1$\times$ 10$^{-4}$ & 1$\times$ 10$^{-4}$ & 1$\times$ 10$^{-9}$ & 1$\times$ 10$^{-9}$ \nl
Region B & 12 & 2277 & 0.005 & 1$\times$ 10$^{-4}$  & 9$\times$ 10$^{-6}$ & 4$\times$ 10$^{-10}$  & 3$\times$ 10$^{-11}$   \nl
Region C & 3 & 958 & 0.003 & 4$\times$ 10$^{-5}$  & 2$\times$ 10$^{-6}$  & 7$\times$ 10$^{-11}$  & 3$\times$ 10$^{-12}$ \nl
\enddata
\end{deluxetable}

\begin{deluxetable}{cccccccc}
\tablenum{8}
\tablewidth{0pt}
\tablecaption{Variable candidates in Phoenix}
\tablehead{
\colhead{N} &    \colhead{X} & \colhead{Y} &
\colhead{$V$} & \colhead{$I$}  &
\colhead{$(V-I)$} & \colhead{$M_{I}$} & \colhead{$\Delta V$}}

\startdata
3814 & 876.11 & 1378.29 &  21.82 & 20.99  & 0.83 & -2.05 & 0.78 $\pm$ 0.11 \nl
3200 & 872.75 & 1278.09 &  22.09 & 21.37 & 0.72 & -1.67 & 0.71 $\pm$ 0.04 \nl
2697 & 903.61 & 1204.08 &  22.06 & 21.68 & 0.38 & -1.36 & 0.32 $\pm$ 0.08  \nl
2981 & 693.51 & 1232.80 &  21.47 & 20.72 & 0.75 & -2.32 & 0.44 $\pm$ 0.03  \nl
\enddata
\end{deluxetable}

\newpage

\begin{deluxetable}{ccccccc}
\tablenum{9}
\tablewidth{0pt}
\tablecaption{ Globular cluster candidates}
\tablehead{\colhead{Number} &    
 \colhead{X} &  \colhead{Y} &
 \colhead{$V$} &
  \colhead{$(V-I)$}
 & \colhead{M$_{V}$} & \colhead{FWHM ($\arcsec$)}}

\startdata
$\#$3 &  1258.66 & 1617.80  & 18.66 & 1.42 & -4.39  & 1.3 \nl
 a &  1018.72 & 1352.04   & 18.91 & - & -4.14 & 1.3 \nl
 b &  1383.09 & 1329.97  & 19.03 & 1.54 & -4.02 & 1.8\nl
 c &  1345. 34 & 1190.75  & 18.81 & 1.39 & -4.24 & 1.3\nl

\enddata
\end{deluxetable}

\end{document}